\documentclass[journal=acs, manuscript=article]{achemso}

\usepackage{graphicx}
\usepackage{dcolumn}

\usepackage[T1]{fontenc}
\usepackage[utf8]{inputenc}
\usepackage{mathptmx}
\usepackage{etoolbox}

\usepackage[normalem]{ulem}

\usepackage{bibentry}

\usepackage{siunitx}
\usepackage{mhchem}
\usepackage{float}
\usepackage{xfrac}
\usepackage{hyperref}
\usepackage{xcolor}
\xdefinecolor{tugreen}{RGB}{128, 186, 38}

\newcommand\sref[2]{\hyperref[#1]{\ref*{#1}(#2)}}

\newcommand\sample{FA$_{0.9}$Cs$_{0.1}$PbI$_{2.8}$Br$_{0.2}$}
\newcommand\mueV{$\,\mu$eV}

\usepackage{placeins}



\title{Long-lived exciton coherence\\in mixed-halide perovskite crystals}

\author{Stefan~Grisard}
\email{stefan.grisard@tu-dortmund.de}
\affiliation{Experimentelle Physik 2, Technische Universit\"at Dortmund, 44221 Dortmund, Germany}

\author{Artur~V.~Trifonov}
\affiliation{Experimentelle Physik 2, Technische Universit\"at Dortmund, 44221 Dortmund, Germany}

\author{Ivan~A.~Solovev}
\affiliation{Experimentelle Physik 2, Technische Universit\"at Dortmund, 44221 Dortmund, Germany}
\alsoaffiliation{Institute of Physics, University of Oldenburg, Carl von Ossietzky Strasse 9-11, 26129 Oldenburg, Germany}

\author{Dmitri~R.~Yakovlev}
\affiliation{Experimentelle Physik 2, Technische Universit\"at Dortmund, 44221 Dortmund, Germany}

\author{Oleh Hordiichuk}
\affiliation{Laboratory of Inorganic Chemistry, Department of Chemistry and
Applied Biosciences, ETH Z\"urich, CH-8093 Z\"urich, Switzerland}
\alsoaffiliation{Laboratory for Thin Films and Photovoltaics, Empa-Swiss Federal Laboratories
for Materials Science and Technology, CH-8600 D\"ubendorf, Switzerland}

\author{Maksym~V.~Kovalenko}
\affiliation{Laboratory of Inorganic Chemistry, Department of Chemistry and
Applied Biosciences, ETH Z\"urich, CH-8093 Z\"urich, Switzerland}
\alsoaffiliation{Laboratory for Thin Films and Photovoltaics, Empa-Swiss Federal Laboratories
for Materials Science and Technology, CH-8600 D\"ubendorf, Switzerland}

\author{Manfred~Bayer}
\affiliation{Experimentelle Physik 2, Technische Universit\"at Dortmund, 44221 Dortmund, Germany}

\author{Ilya~A.~Akimov}
\email{ilja.akimov@tu-dortmund.de}
\affiliation{Experimentelle Physik 2, Technische Universit\"at Dortmund, 44221 Dortmund, Germany}

\begin{document}

\date{\today}

\begin{abstract}
    Compositional engineering of the optical properties of hybrid organic-inorganic lead halide perovskites is 
    one of the cornerstones for the realization of efficient solar cells and tailored light-emitting devices. 
    We study the effect of compositional disorder on coherent exciton dynamics in 
    a mixed FA$_{0.9}$Cs$_{0.1}$PbI$_{2.8}$Br$_{0.2}$ perovskite crystal using photon echo spectroscopy. 
    We reveal that the homogeneous linewidth of excitons can be as narrow as 16\mueV at a temperature of 1.5\,K.
    The corresponding exciton coherence time of $T_2=83$\,ps is exceptionally long being attributed to the localization of excitons 
    due to variation of composition at the scale of ten to hundreds of nanometers. 
    From spectral and temperature dependences of the two- and three-pulse photon echo decay we conclude that for low-energy excitons, 
    pure decoherence associated with elastic scattering on phonons is comparable with the exciton lifetime, while for excitons with 
    higher energies, inelastic scattering to lower energy states via phonon emission dominates. 
\end{abstract}

\maketitle

Formamidinium lead tri-iodide (\ce{FAPbI_3}) represents a promising candidate for novel technologies in the field of 
photovoltaics and light-emitting devices~\cite{min_efficient_2019, chen_advances_2021, zheng_development_2022, docampo_long-term_2016}. 
It is therefore of high relevance to gain a clear understanding of the light emission and absorption properties of the material and to tune 
them for optimal application design. The ability to engineer the band gap and improve the poor phase stability of 
pure \ce{FAPbI_3} by compositional variation has raised strong interest in 
the \ce{FA_{1-x}Cs_{x}PbI_{3-y}Br_y} compounds~\cite{HU_tuning, mcmeekin_mixed-cation_2016}. Here, partial replacement of the 
large \ce{FA^+} cation by smaller \ce{Cs^+} stabilizes the crystal lattice whereas changing the 
Bromine/Iodine ratio allows tuning the band gap energy in a range 
between \SI{1.5}{\eV} and \SI{2.5}{\eV}~\cite{xue_localized_2018, prasanna_band_2017, mcmeekin_mixed-cation_2016}. 
Although compositional substitution has proven to offer high flexibility to engineer the optoelectronic properties of hybrid 
organic-inorganic lead halide perovskites, an understanding of the effect of such substitutions on the energy structure of 
the material remains elusive. In particular, the role of spatial disorder, which strongly affects coherent exciton properties, is not 
well understood. 

A powerful approach to reveal the effect of compositional substitution on the energy structure of semiconductors in the vicinity of the band gap 
is based on the analysis of 
the homogeneous linewidth of excitons at cryogenic temperatures~\cite{noll_picosecond_1990, poltavtsev_long_2018}. 
The homogeneous exciton linewidth is influenced by various factors, including exciton-phonon interaction and the degree of exciton localization. 
As the exciton line is usually inhomogeneously broadened even at cryogenic temperatures, non-linear optical techniques based on photon echoes or 
two-dimensional Fourier spectroscopy have to be used to access the homogeneous linewidth~\cite{cundiff_coherent_2008}. 
For lead halide perovskite semiconductors, coherent optical techniques have been applied to address exciton physics and to identify biexciton 
formation in bulk \ce{MAPbI3}~\cite{trifonov_photon_2022}, size-dependent line broadening mechanisms in \ce{CsPbI_3} 
nanoplatelets~\cite{liu_toward_2021}, and exciton fine structure in~\ce{CsPbI_3} nanocrystals~\cite{liu_multidimensional}. Moreover, 
perovskite nanocrystals exhibit highly efficient single-photon
emission~\cite{utzat_coherent} and, being self-organized into three-dimensional superlattices, show superfluorescence, which results from the coherent 
coupling of individual uncorrelated emitters~\cite{raino_superfluorescence_2018}. These reports
generated excitement about possible applications of perovskite semiconductors as bright quantum light sources, e.g., 
in quantum information processing, expanding the range of potential applications beyond photovoltaics and optoelectronics.
So far, the coherence times in bulk perovskites were reported to be short (few picoseconds in MAPbI$_3$ \cite{trifonov_photon_2022, nazarov_photon_2022}) 
as expected for free and weakly localized excitons.

In this paper, we apply transient photon echo spectroscopy to a~\sample\, single crystal at a 
temperature of $T = \SI{1.5}{\kelvin}$. We observe exceptionally long coherence times of excitons 
of \SI{83}{\pico\second}, corresponding to a homogeneous exciton linewidth of~\SI{16}{}\mueV, which is three orders of magnitude smaller than the 
inhomogeneous broadening and at least one order of magnitude narrower as compared to the homogeneous linewidth observed in similar ternary 
\ce{MAPbI3} perovskite crystals. The combination of the two- and three-pulse photon echo techniques 
allows us to quantify the contribution of elastic scattering events, i.e. pure decoherence effects, 
to the broadening of the exciton line. By decomposing the linewidth into contributions of pure decoherence and energy 
relaxation as a function of photon energy and temperature, we reveal clear signatures of exciton localization
that resemble the behavior of zero-dimensional systems with discrete energy spectrum, e.g. 
semiconductor quantum dots~\cite{bayer_bridging_2019}. 
Our results demonstrate that lead halide perovskite semiconductors may have exceptionally long exciton coherence times being comparable with 
those in perovskite nanocrystals. 
This provides an ideal playground for investigations of many-body interactions in dense exciton ensembles~\cite{chemla_many-body_2001}
and coherent collective effects, which are absent in strongly isolated semiconductor nanocrystals.

\begin{figure*}
    \centering
    \includegraphics[scale = 1]{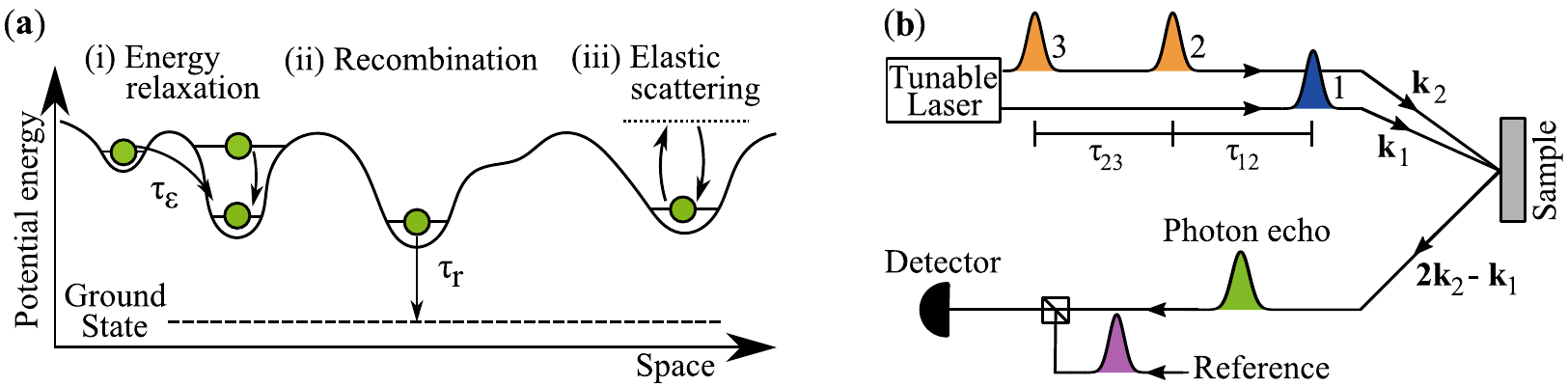}
    \caption{(a) Visualization of possible contributions to the exciton linewidth in the context of localized excitons 
    in a random potential. The population decay time $T_1$ involves 
    contributions of (i) energy relaxation and (ii) recombination. Elastic scattering events
    as shown in (iii) represent a pure decoherence mechanism. (b) Schematic drawing of the experimental setup for transient 
    four-wave-mixing spectroscopy.}
    \label{fig: fig01}
\end{figure*}  

{\bf Experimental technique}. 
The studied \sample\ single crystals
were grown by a modification of a previously described method~\cite{nazarenko_single_2017}. 
The details of sample preparation are given in the Supporting Information (section S1).
Our spectroscopic technique is based on the three-pulse photon echo (3PE). The 3PE results from the optical four-wave-mixing (FWM) response 
with signal field $E_S \propto E_1^*E_2E_3$ of an 
inhomogeneously broadened ensemble of emitters after resonant excitation with three temporally separated laser pulses with electric 
field amplitudes $E_i$ and corresponding complex conjugated $E_i^*$. Throughout the paper, all pulses are linearly co-polarized. 
The photon echo appears at time $t=2\tau_{12}+\tau_{23}$, where $\tau_{ij}$ is the temporal delay between pulses $E_i$ and $E_j$. 
The amplitude of the 3PE pulse depends on the two-time delays as
\begin{equation}
    E_S \propto \exp\left(- \frac{2\tau_{12}}{T_2} - \frac{\tau_{23}}{T_1}\right), 
    \label{eq: SPE_decay}
\end{equation} 
where $T_2$ is the exciton coherence time and $T_1$ is the decay time of the population at a 
specific energy. Thus, a measurement of the amplitude of the 3PE pulse as a function of $\tau_{12}$ or $\tau_{23}$ allows us to 
access $T_2$ and $T_1$ independently. For a more detailed description of the 3PE technique we refer to the Supporting 
Information (section S2) and Ref.~\citenum{cundiff_coherent_2008}.

The population decay time $T_1$ includes contributions of energy relaxation and recombination of excitons, which we 
visualize in Figure~\sref{fig: fig01}{a} within the context of localized excitons in a spatially modulated potential energy landscape. 
The energy relaxation on a timescale $\tau_\varepsilon$, case (i) in Figure~\sref{fig: fig01}{a}, may result from the tunneling of excitons 
between localization sites of different energy or the relaxation to lower energy levels within a localization site. For excitons in the 
lower energy states, the population decay time is limited by the 
finite lifetime through the exciton recombination on a timescale $\tau_r$, which we show in case (ii). 
The exciton population decay time can be written as 
\begin{equation}
    \frac{1}{T_1} = \frac{1}{\tau_\varepsilon} + \frac{1}{\tau_r}.
\end{equation} 

In this study, we use only the first two pulses to measure $T_2$, leading to the two-pulse photon echo (2PE), 
which is equivalent to the case $\tau_{23} = 0$ in Equation~\eqref{eq: SPE_decay}. 
Combining the experimental results for $T_1$ and $T_2$ enables us to quantify the contribution of pure decoherence mechanisms, i.e., mechanisms that 
leave the exciton population unaffected while changing the phase of the excited state's wave function. 
Such processes involve, for example, elastic scattering events with phonons, as we show in case (iii) in Figure~\sref{fig: fig01}{a}. 
Here, the exciton is excited to and deexcited from a virtual state through a two-phonon interaction, which leaves the energy unaffected, but changes 
the phase of the exciton wave function. In the absence of pure decoherence mechanisms, the coherence time is twice as long as the population decay time. 
The coherence time $T_2$ is therefore composed as
\begin{equation} 
    \frac{1}{T_2} = \frac{1}{2T_1} + \frac{1}{T_2'}  
    \label{eq: pure_T2}
\end{equation}
with the pure decoherence time $T_2'$. 

The experimental scheme for time-resolved four-wave-mixing spectroscopy is illustrated in Figure~\sref{fig: fig01}{b}. 
The temperature of the \sample\, crystal is set to \SI{1.5}{\kelvin} unless stated otherwise. Laser pulses with a duration of 
\SI{4.6}{\pico\second} (\SI{0.4}{\milli\eV} spectral width) and a repetition rate of \SI{75.75}{\mega\hertz} are tuned in resonance 
with the exciton transition ($\hbar\omega \approx \SI{1.52}{\eV}$). As depicted in Figure~\sref{fig: fig01}{b}, the second and third pulses 
impinge on the sample sharing the same wavevector $\mathbf{k}_3 = \mathbf{k}_2 \neq \mathbf{k}_1$. In this way, the 2PE response 
with $E_S \propto E_1^*(E_2)^2$, formed by the first and second pulse is emitted in the same phasematched direction $2\mathbf{k}_2 - \mathbf{k}_1$
as the 3PE and can be used for the measurement 
of the coherence time $T_2$. The photon echo signal is resolved in time using the cross-correlation with a reference pulse that is delayed by 
$\tau_\mathrm{ref}$ relative to the first pulse. 
More details on the experimental technique are given in the Supporting Information (section S2) and Ref.~\citenum{poltavtsev_photon_2018}.

\begin{figure*}
    \centering
    \includegraphics[width = \textwidth]{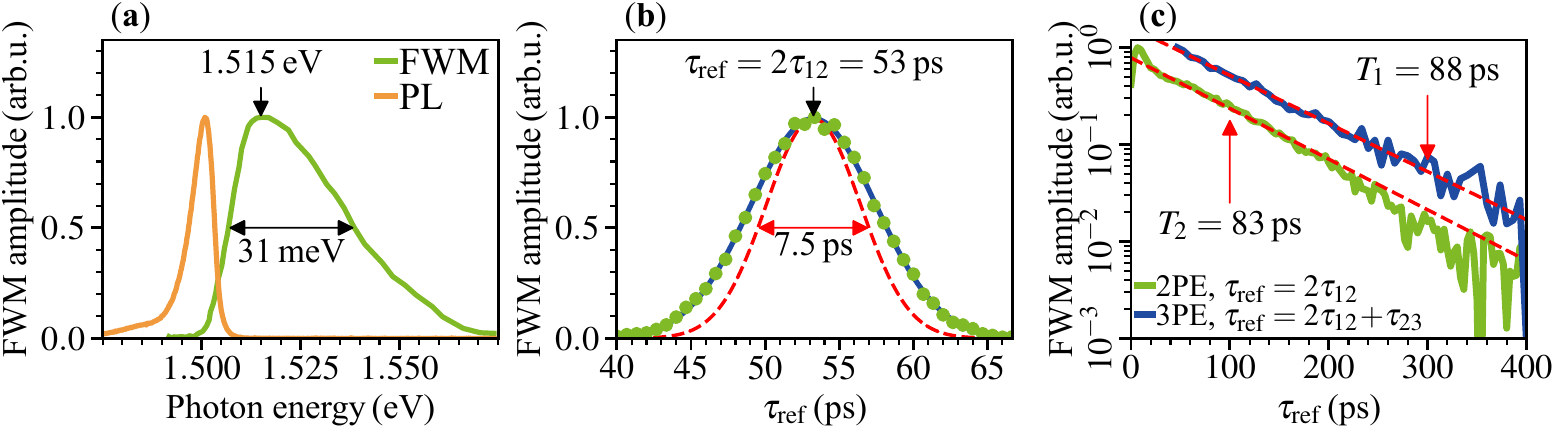}
    \caption{(a) Four-wave-mixing spectrum of the \sample\, crystal. Arrows indicate the peak position and FWHM. For 
    comparison, we show the photoluminescence (PL) spectrum. 
    (b) Time-resolved photon echo pulse, measured for $\tau_{12} = \SI{26.5}{\pico\second}$. The blue line is a fit to a 
    Gaussian function, the red line is the calculated deconvolution from the reference pulse envelope. 
    (c) Decays of the two- and three-pulse photon echo. Both decays are measured at a photon energy 
    of \SI{1.515}{\eV} at $T = \SI{1.5}{\kelvin}$. The dashed red lines represent fits to single exponential 
    functions.}
    \label{fig: fig02}
\end{figure*}

{\bf Coherent dynamics of excitons at low temperature}. 
As a first step, we analyze the four-wave-mixing spectrum of the sample close to the band gap 
energy \SI{1.52}{\eV} as reported in Ref.~\citenum{kirstein_lead-dominated_2022} for a crystal of the same composition. 
For this purpose, we measure the four-wave-mixing signal arising from the two-pulse sequence for $\tau_{12} = 0$ as a function of the photon energy of the 
spectrally narrow picosecond laser pulses. The resulting spectral dependence is shown in Figure~\sref{fig: fig02}{a}. We find an asymmetric lineshape 
with a peak at \SI{1.515}{\eV}. The low energy part of the spectrum has a half width at half maximum (HWHM) of \SI{8}{\milli\eV}, while the high energy part is 
significantly broader with HWHM \SI{23}{\milli\eV}. The total full width at half maximum (FWHM) amounts to \SI{31}{\milli\eV} 
as indicated in Figure~\sref{fig: fig02}{a}. The width on the low energy side coincides with the result of photoluminescence (PL) 
excitation spectroscopy performed in Ref.~\citenum{kirstein_lead-dominated_2022}, whereas the broad high energy tail may be influenced by contributions from 
excited exciton states. 
The large width as well as the asymmetry of the spectrum hint at a dominant contribution of inhomogeneous broadening~\cite{moody_exciton-exciton_2011}, 
which corresponds to about 16~meV as determined by the low energy side. The orange line in 
Figure~\sref{fig: fig02}{a} shows the corresponding time integrated PL spectrum. 
The PL spectrum is shifted by \SI{14}{\milli\eV} to lower energies with respect to the maximum of the 
FWM spectrum. As revealed by a time-resolved study in Ref.~\citenum{kirstein_lead-dominated_2022}, 
the PL spectrum is dominated by slow recombination dynamics of spatially separated electrons and holes.

The presence of inhomogeneous broadening of the exciton line can be confirmed by introducing a delay of $\tau_{12} = \SI{26.5}{\pico\second}$ between 
the first and second pulse with a photon energy of \SI{1.515}{\eV}. In this case, as presented in Figure~\sref{fig: fig02}{b}, the FWM response of the sample 
is given by a photon echo pulse. The pulse is well described by a Gaussian centered at $\tau_\mathrm{ref} = 2\tau_{12} = \SI{53}{\pico\second}$ with a 
FWHM of \SI{9}{\pico\second}. A deconvolution of the fitted Gaussian from the temporal profile of the reference pulse is shown by a
red curve in Figure~\sref{fig: fig02}{b} revealing a FWHM of \SI{7.5}{\pico\second} of the echo pulse. Note that the laser spectrum with a width of 
\SI{0.4}{\milli\eV} is more narrow 
than the inhomogeneously broadened exciton resonance. Therefore, the duration of the photon echo pulse is approximately given by the duration of the laser 
pulses (\SI{6.5}{\pico\second} associated with the FWHM of the electric field envelope). Note that the use of spectrally narrow picosecond pulses allows us 
to gain information about the homogeneous exciton linewidth with spectral resolution over the inhomogeneous broadening, which we will discuss in more detail below. 

In Figure~\sref{fig: fig02}{c}, we present the result of a $T_2$ measurement at a photon energy of \SI{1.515}{eV} using the 2PE. Here, the photon echo 
amplitude at $\tau_\text{ref} = 2\tau_{12}$ is recorded as a function of $\tau_\mathrm{ref}$. The 2PE amplitude follows a single exponential 
decay $\propto\exp(2\tau_{12} / T_2)$ with $T_2 = \SI{83(1)}{\pico\second}$ corresponding to a narrow Lorentzian homogeneous line with 
FWHM of $\Gamma_2 = 2\hbar/T_2 = \SI{15.9(2)}{}$\mueV. Therefore, we conclude that the spectrum shown in 
Figure~\sref{fig: fig02}{a} is dominantly inhomogeneously broadened. In the studied sample, the inhomogeneous broadening exceeds the homogeneous broadening by three 
orders of magnitude. It is remarkable that the coherence time of excitons in the~\sample \,crystal is significantly longer as compared to the 
measured values in the ternary perovskite crystal \ce{MAPbI_3} ($T_2 = \SI{0.8}{\pico\second}$ in Ref.~\citenum{trifonov_photon_2022} and 
\SI{8}{\pico\second} in Ref.~\citenum{nazarov_photon_2022}). 
The observed value in our sample rather resembles results obtained in low-dimensional systems such as 
\ce{CsPbBr2Cl} nanocrystals ($T_2\approx\SI{25}{\pico\second}$ in Ref.~\citenum{becker_long_2018}). 
The long coherence time suggests a reduced interaction of excitons with their environment, which we attribute to the localization of excitons in potentials formed 
by spatial fluctuations of composition.  

Coherent laser spectroscopy was successfully used in the past to study various semiconductor systems that are subject to compositional or geometric 
disorder, such as GaAs/(Al,Ga)As quantum wells that exhibit spatial fluctuations of the well width~\cite{cundiff_coherent_1992}. 
Spatial disorder can lead to exciton localization and, consequently, to a reduction in their mobility accompanied by a narrowing of the homogeneous linewidth. 
Localized excitons in mixed semiconductor crystals were studied in a variety of semiconductors such as II-VI 
alloys \ce{CdS_xSe_{1-x}}~\cite{permogorov_emission_1981, noll_picosecond_1990} or \ce{ZnSe_{1-x}Te_x}~\cite{stolz_selective_1988, permogorov_localisation_1989}.
The formation of quantized energy levels due to exciton localization in three dimensions can lead to a drastic reduction of 
phonon-assisted energy relaxation~\cite{uskov_dephasing_2000, benisty_intrinsic_1991, yu_time_1996, bardot_exciton_2005, takagahara_theory_1999}. 
Further, purely elastic scattering events, that involve two phonons can be enhanced by the localization~\cite{fan_pure_1998, borri_exciton_2005, moody_exciton-exciton_2011}. 
Therefore, as a next step of our in-depth study, we decompose the coherence time $T_2$ into contributions arising from population decay and purely elastic scattering. 
For that, we measure the decay of the amplitude of the 3PE  
centered at $\tau_\text{ref} = 2\tau_{12} + \tau_{23}$ for a fixed value of~$\tau_{12} = \SI{20}{\pico\second}$. 
The 3PE amplitude shown in Figure~\sref{fig: fig02}{c} by a blue line decays exponentially with $T_1 = \SI{88(3)}{\pico\second}$. 
The measured value for $T_2 = \SI{83(1)}{\pico\second}$  
at a photon energy of~\SI{1.515}{\eV} is noticeably smaller than $2T_1 \approx \SI{176}{\pico\second}$ (as expected in lifetime limited case $T_2 = 2T_1$), 
which demonstrates a significant contribution of pure decoherence mechanisms. Using Equation~\eqref{eq: pure_T2}, we obtain a value of $T_2' = \SI{158(6)}{\pico\second}$ 
and the contribution to the homogeneous linewidth $\Gamma_2'=2\hbar/T_2'= \SI{8.3(3)}{}$\mueV. 

\begin{figure*}
    \centering
    \includegraphics[scale = 1]{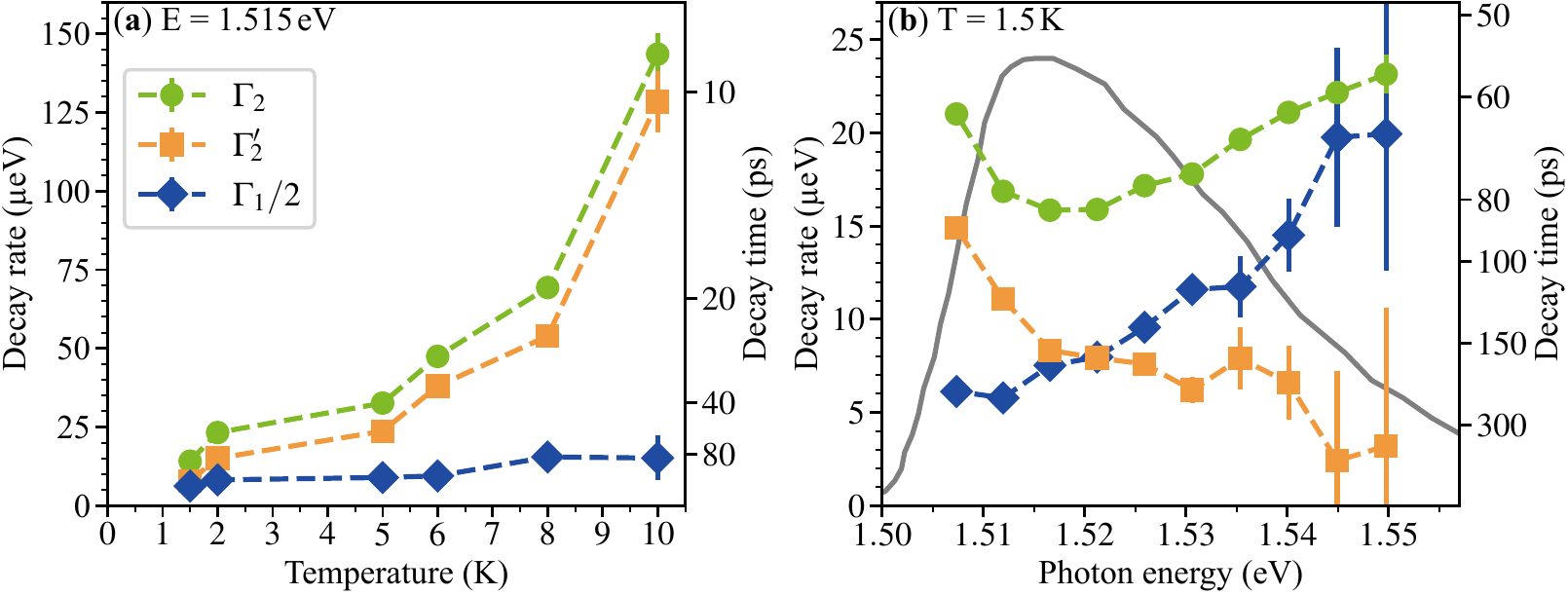}
    \caption{(a) Contributions to the homogeneous exciton linewidth $\Gamma_2 = 2\hbar/T_2$ associated with the population decay 
    rate $\Gamma_1/2=\hbar/T_1$ and pure decoherence rate $\Gamma_2'=2\hbar/T_2'$ as a function of temperature, 
    measured at a photon energy of \SI{1.515}{\eV}. Here, $\Gamma_2$ and $\Gamma_1$ result from the measurement of the photon echo decay as a function of 
    $\tau_{12}$ and $\tau_{23}$, respectively, whereas $\Gamma_2'$ is calculated using Equation~\eqref{eq: pure_T2}. 
    (b) Dependence of $\Gamma_2$, $\Gamma_1 / 2$, and $\Gamma_2'$ on 
    the central photon energy of the ps-laser pulses, measured at a temperature of \SI{1.5}{\kelvin}. 
    The solid line shows the FWM spectrum for comparison in arbitrary units.}
    \label{fig: fig03}
\end{figure*}    

{\bf Temperature and spectral depependences}.
To gain insight into the origin of pure decoherence mechanisms, we perform a temperature-dependent measurement. 
Figure~\sref{fig: fig03}{a} shows two contributions arising from pure decoherence rate $\Gamma_2'$ and population decay 
rate $\Gamma_1/2 = \hbar / T_1$ as a function of temperature. In this representation, the sum of these quantities determines the homogeneous 
linewidth of exciton $\Gamma_2 = 2 \hbar / T_2$ (Equation~\eqref{eq: pure_T2}). The right axis in Figure~\sref{fig: fig03}{a} shows the corresponding 
inverted scale for decay times. We observe that the pure decoherence rate $\Gamma_2'$
grows faster compared to $\Gamma_1/2$ in a non-linear manner from $\SI{8}{}$\mueV\, at \SI{1.5}{\kelvin} to \SI{130}{}\mueV\, at \SI{10}{\kelvin}. 
$\Gamma_1/2$ in turn does not experience a significant rise and remains below \SI{20}{}\mueV\, in the observed temperature 
range.  
This observation is in agreement with the localization of excitons, since the discrete energy spectrum of the localized excitons suppresses 
their energy relaxation, and in particular activation of higher energy states via phonon absorption. Such processes would lead to a 
rise of $\Gamma_1$ with increasing temperature~\cite{bardot_exciton_2005, yu_time_1996}. 
Pure elastic scattering events including two phonons in turn are more important for localized excitons due to the enhanced spread of the exciton wavefunction 
in momentum space~\cite{fan_pure_1998}. In this way, 
pure decoherence is enhanced by the increasing phonon population with temperature. 

To address exciton states with different energies, we measure a spectral dependence of the exciton linewidth. Within an ensemble of excitons localized by compositional 
fluctuations, excitons contributing at the low energy side of the spectrum are 
stronger localized compared to excitons at the high energy side. In Figure~\sref{fig: fig03}{b} we plot results for $\Gamma_1$ and $\Gamma_2'$ as a function of the 
laser photon energy, measured at a temperature of \SI{1.5}{\kelvin}. The FWM spectrum is shown in the same graph for comparison. We observe, that the contribution of 
population decay to the homogeneous linewidth $\Gamma_1 / 2$ monotonically increases towards higher energies from roughly \SI{5}{}\mueV\, to \SI{20}{}\mueV, 
while the pure decoherence rate $\Gamma_2'$ 
gradually decreases from \SI{15}{}\mueV\, to \SI{3}{}\mueV\, and becomes significantly smaller than $\Gamma_1/2$. 
Such behavior leads toward the limiting case at higher energies where the homogeneous 
linewidth is determined by the population decay, i.e. $\Gamma_2 = \Gamma_1 / 2 \approx \SI{23}{}$\mueV. 

These observations are in agreement with the aforementioned correlation between the efficiency of energy relaxation and pure decoherence mechanisms 
for excitons with different degree of localization. The decrease of the pure decoherence rate $\Gamma_2'$ with increasing photon energy 
demonstrates that excitons with higher energies possess a smaller degree of localization because 
the elastic multiple-phonon scattering mechanism (case (iii) in Figure~\sref{fig: fig01}{a}) is significantly reduced. 
In contrast, the energy relaxation rate $1/\tau_{\epsilon}$ is negligible for excitons in low-energy states due to the 
absence of lower energy states that can be accessed via phonon emission~\cite{bockelmann_exciton_1993, benisty_intrinsic_1991}.
Therefore, the population decay time measured on the low energy side of the spectrum 
at \SI{1.510}{\eV}, $T_1 = \SI{108(4)}{\pico\second}$, is the best estimate for the exciton lifetime $\tau_r$, which takes place due to exciton
recombination (case (ii) in Fig.~\sref{fig: fig01}{a}). At higher energy, the role of confined exciton states with larger quantum number as well as hopping of 
weakly localized excitons into the deeper potential sites accompanied by phonon emission starts to play an important role, i.e. $1/\tau_{\epsilon}$ 
dominates the population decay rate (case (i) in Figure~\sref{fig: fig01}{a}). 

We stress that under resonant excitation with a photon energy of \SI{1.515}{\eV}, we do not observe the activation 
of higher energy states up to a temperature of \SI{10}{\kelvin}, which would result in a significant rise of  
$\Gamma_1$ in Figure~\sref{fig: fig03}{b}. The population decay is therefore limited by $\tau_r$. 
This means that the energy distance to the next higher-lying confinement state 
is at least larger than $k_\mathrm{B}T=\SI{1}{\milli\eV}$, where $k_\mathrm{B}$ is the Boltzmann constant. 
Using this value as the minimal energy spacing $\Delta_{\rm min}$ between the two lowest states of excitons 
confined in a spherical potential, we can estimate the maximum diameter of the potential 
trap $d_\mathrm{max}$. Taking into account that the potential height $U\approx\SI{16}{\milli\eV}$, as follows from the inhomogeneous broadening of the 
FWM spectrum in Figure~\sref{fig: fig02}{a}, is much larger than $\Delta_{\rm min}$, 
we obtain $d_{max} = \pi\hbar\sqrt{6/(\Delta_{\rm min} m_X)} \approx \SI{95}{\nano\meter}$, 
where $m_X=0.5m_0$ is the translational exciton mass taken from Refs.~\citenum{galkowski_determination_2016, yang_impact_2017, baranowski_exciton_2020}. 
The minimum size can be estimated as $d_{\rm min}=\pi\hbar/\sqrt{2 U m_X}\approx \SI{7}{\nano\meter}$~\cite{klingshirn_semiconductor_2007}. 
Therefore, we conclude that potential fluctuations with a magnitude of about \SI{16}{\milli\eV} 
take place in the \sample\, crystal on a scale of \SI{7}{\nano\meter} to \SI{95}{\nano\meter}.

\FloatBarrier
In conclusion, we demonstrated an exceptionally long exciton coherence time of \SI{83}{\pico\second} in the mixed perovskite single 
crystal~\sample\, using photon echo spectroscopy. The coherence time is comparable to the observed exciton lifetime 
of \SI{88}{\pico\second} measured at \SI{1.515}{\eV}. We associate these findings to the formation of potential traps by random 
fluctuations of the cation and anion content in the crystal with a magnitude of about \SI{16}{\milli\eV} 
and a spatial extent of tens to hundreds of nanometers. 
Since potential fluctuations directly control the charge transport, as well as the absorption and emission properties 
of the material, we propose to extend the presented study for various compositions. This will offer valuable information for 
optimal design of optoelectronic devices based on \ce{FA_xCs_{1-x}PbI_{3-y}Br_y}. Furthermore, long coherence times, which are comparable to the 
exciton lifetime, are important for the application of perovskite semiconductors as bright quantum light sources, e.g., in quantum information processing, expanding 
the range of potential applications beyond photovoltaics and optoelectronics.

\section{Acknowledgements}

The authors acknowledge financial support by the Deutsche Forschungsgemeinschaft via the SPP2196 Priority Program (Project AK 40/13-1)
as well as through the Collaborative
Research Center TRR 142/3 (Grant No. 231447078, Project No. A02).
The work at ETH Zurich was financially supported by the Swiss National Science 
Foundation (grant agreement 186406, funded in conjunction with SPP219 through the DFG-SNSF bilateral program) 
and by ETH Zurich through ETH+ Project SynMatLab.
I.~A.~S. acknowledges support by TU Dortmund core funds.

\providecommand{\latin}[1]{#1}
\makeatletter
\providecommand{\doi}
  {\begingroup\let\do\@makeother\dospecials
  \catcode`\{=1 \catcode`\}=2 \doi@aux}
\providecommand{\doi@aux}[1]{\endgroup\texttt{#1}}
\makeatother
\providecommand*\mcitethebibliography{\thebibliography}
\csname @ifundefined\endcsname{endmcitethebibliography}
  {\let\endmcitethebibliography\endthebibliography}{}

\def\thesection{S\arabic{section}}
\def\theequation{S\arabic{equation}}
\def\thefigure{S\arabic{figure}}
\def\thetable{S\arabic{table}}
\setcounter{equation}{0}

\section{Supplementary Information}

\textbf{S1 Samples}. Lead(II) iodide (\ce{PbI2}, \SI{99}{\percent}) and $\gamma$-butyrolactone (GBL, $\leq$ \SI{99}{\percent}) 
were purchased from Sigma-Aldrich. Lead(II) bromide (\ce{PbBr2}, 98+\%) was purchased from Acros. 
Cesium iodide (\ce{CsI}, \SI{99.9}{\percent}), formamidine acetate (FA(OAc), \SI{99}{\percent}) and hydroiodic acid (\SI{57}{\percent}, 
stabilized with \SI{1.5}{\percent} hypophosphorous acid) were purchased from ABCR. All chemicals were used as received without further purification. 

\sample single crystals were grown by a modification of the method that was previously used to synthesize formamidinium iodide (FAI) crystals 
in Ref.~1. 
First, a stoichiometric solution of \ce{CsI}, FAI, \ce{PbI2}, and \ce{PbBr2} 
in $\gamma$-butyrolactone (GBL) as solvent was prepared (\SI{0.9}{M} w.r.t. Pb) by 
dissolving the powders of the compounds. The solution was filtered 
through 0.2\,$\mu$m polytetrafluoroethylene (PTFE) syringe filter. About \SI{6}{\milli\liter} of the precursor solution was 
filled into a \SI{20}{\milli\liter} PTFE-capped glass vials and placed into a glycerol bath on a hot plate. This solution was then 
slowly (over 3--4 hours) heated until crystals appeared and grew to several millimeter size (heating to ca. 100--\SI{110}{\celsius}). 
Several of the crystals were quickly transferred with tweezers to fresh initial solutions 
preheated to 80--\SI{90}{\celsius}. 
The crystals (seeds) were allowed to dissolve partially. After that, heating was resumed and continued 
slowly (4--5 hours) to 115--\SI{120}{\celsius}. The crystals were removed with tweezers, dried with soft tissues, and transferred to a 
\ce{N2}-filled glovebox. 

\textbf{S2 Transient four-wave-mixing}. Non-linear optical techniques, such as photon echo spectroscopy, provide a powerful tool to overcome the limitations of inhomogeneous broadening and to 
study the homogeneous linewidth of excitons in semiconductors. 
In the four-wave-mixing process, the first pulse $E_1$ resonantly excites a coherent superposition between ground and exciton 
state that evolves in time according to $\propto \exp(-i \omega t - t / T_2)$. Here, $\omega$ is the transition frequency of the 
exciton ($\hbar\omega \approx \SI{1.52}{\eV}$ in our case) and $T_2$ is the exciton coherence time that defines the 
homogeneous exciton linewidth. Within an inhomogeneous ensemble of excitons, where the transition frequency $\omega$ is 
statistically distributed, the macroscopic response of the system dephases quickly within the dephasing time $T_2^*$ and, therefore, 
hinders measuring the homogeneous exciton coherence time $T_2$. After a time delay $\tau_{12}$, the second pulse $E_2$ creates a spectral 
grating of the exciton population that decays exponentially with the population decay time $T_1$. The third pulse $E_3$ creates 
again a coherence for each individual exciton whose time evolution $\propto \exp(i \omega t - t / T_2)$ is conjugated 
with respect to the time interval between the first and second pulse. Therefore, a macroscopic 
polarization is rebuilt that results in the emission of a three-pulse photon echo (3PE) at time $\tau_{12}$ after the third pulse in the 
phase-matched direction $\mathbf{k}_3 + \mathbf{k}_2 - \mathbf{k}_1$ (compare Figure~1(b) in the main text), 
with the wavevectors $\mathbf{k}_i$ of the respective laser fields. 
The overall dependence of the 3PE amplitude on the time delays is thus given by 
\begin{equation}
    E_S \propto \exp\left(- \frac{2\tau_{12}}{T_2} - \frac{\tau_{23}}{T_1}\right). 
    \label{eq: SPE_decay_Methods}
\end{equation} 
A measurement of the amplitude of the 3PE pulse as a function of $\tau_{12}$ or $\tau_{23}$ allows to access $T_1$ and $T_2$ independently. 

The \sample\, crystal is placed in a variable temperature insert of a helium bath cryostat and cooled 
down to a \SI{1.5}{\kelvin}. We use a tunable Ti:Sa laser as a source of optical pulses with a duration of 
\SI{4.6}{\pico\second} (\SI{0.4}{\milli\eV} spectral width) and a repetition rate of \SI{75.75}{\mega\hertz}. 
Each pulse is split into three pulses whose relative 
delays $\tau_{12}$ and $\tau_{23}$ are adjusted using mechanical delay lines. The second and third pulses impinge on the sample sharing the same 
wavevector $\mathbf{k}_3 = \mathbf{k}_2 \neq \mathbf{k}_1$ (here, $\mathbf{k}_1$ and $\mathbf{k}_2$ correspond to incident angles of 
\SI{3}{\degree} and \SI{4}{\degree}, respectively). In this way, the two-pulse photon echo (2PE) response with $E_S \propto E_1^*(E_2)^2$, formed 
by the first and second pulse is emitted in the same direction as the 3PE and can be used for the measurement of the coherence 
time $T_2$. The photon echo is detected using the heterodyne detection where we interfere the weak signal field with a strong reference pulse 
produced by the same laser and record the signal using a photodiode connected to a lock-in amplifier. Temporal resolution is acquired by changing the 
arrival time $\tau_\mathrm{ref}$ of the reference pulse on the photodiode with respect to the first pulse. Details on the underlying modulation technique are 
described in Ref.~2. 

\textbf{References}

(1) Nazarenko, O.; Yakunin, S.; Morad, V.; Cherniukh, I.; Kovalenko, M. V. Single crystals
of caesium formamidinium lead halide perovskites: solution growth and gamma dosimetry.
NPG Asia Mater 2017, 9, e373--e373.

(2) Poltavtsev, S. V.; Yugova, I. A.; Akimov, I. A.; Yakovlev, D. R.; Bayer, M. Photon Echo
from Localized Excitons in Semiconductor Nanostructures. Phys. Solid State 2018, 60,
1635--1644.

\end{document}